\documentclass[sigconf]{acmart}
\graphicspath{ {./images/} }
\usepackage{subcaption}
\DeclareMathOperator*{\argmax}{arg\,max}
\renewcommand\footnotetextcopyrightpermission[1]{} 
\AtBeginDocument{%
  \providecommand\BibTeX{{%
    \normalfont B\kern-0.5em{\scshape i\kern-0.25em b}\kern-0.8em\TeX}}}

\acmConference[PASC '23]{PASC '23}{June 26--29, 2023}{Davos, Switzerland}
\begin{document}

\title{Causal Discovery and Optimal Experimental Design for Genome-Scale Biological Network Recovery}

\author{Ashka Shah}
\affiliation{%
  \institution{University of Chicago}
  \city{Chicago}
  \state{IL}
    \country{USA}
}
  \email{shahashka@uchicago.edu}

\author{Arvind Ramanathan}
\affiliation{%
  \institution{Argonne National Laboratory}
  \city{Lemont}
  \state{IL}
    \country{USA}
  }

\email{ramanathana@anl.gov}

\author{Valerie Hayot-Sasson}
\affiliation{%
  \institution{University of Chicago }
  \city{Chicago}
  \state{IL}
  \country{USA}
  }
\email{vhayot@uchicago.edu}

\author{Rick Stevens}
\affiliation{%
  \institution{University of Chicago }
  \city{Chicago}
  \state{IL}
  \country{USA}
  }
\email{stevens@cs.uchicago.edu}
\begin{abstract}
Causal discovery of genome-scale networks is important for identifying pathways from genes to observable traits --e.g. differences in cell function, disease, drug resistance and others. Causal learners based on graphical models rely on interventional samples to orient edges in the network. However, these models have not been shown to scale up the size of the genome, which are on the order of $10^3$-$10^4$ genes. We introduce a new learner, SP-GIES, that jointly learns from interventional and observational datasets and achieves almost 4x speedup against an existing learner for 1,000 node networks. SP-GIES achieves an AUC-PR score of 0.91 on 1,000 node networks, and scales up to 2,000 node networks -- this is 4x larger than existing works. We also show how SP-GIES improves downstream optimal experimental design strategies for selecting interventional experiments to perform on the system. This is an important step forward in realizing causal discovery at scale via autonomous experimental design.
\end{abstract}

\keywords{causality, structure learning, optimal experimental design, genotype-phenotype mapping}

\maketitle

\section{Introduction}
\begin{figure*}
\includegraphics[width=0.8\linewidth]{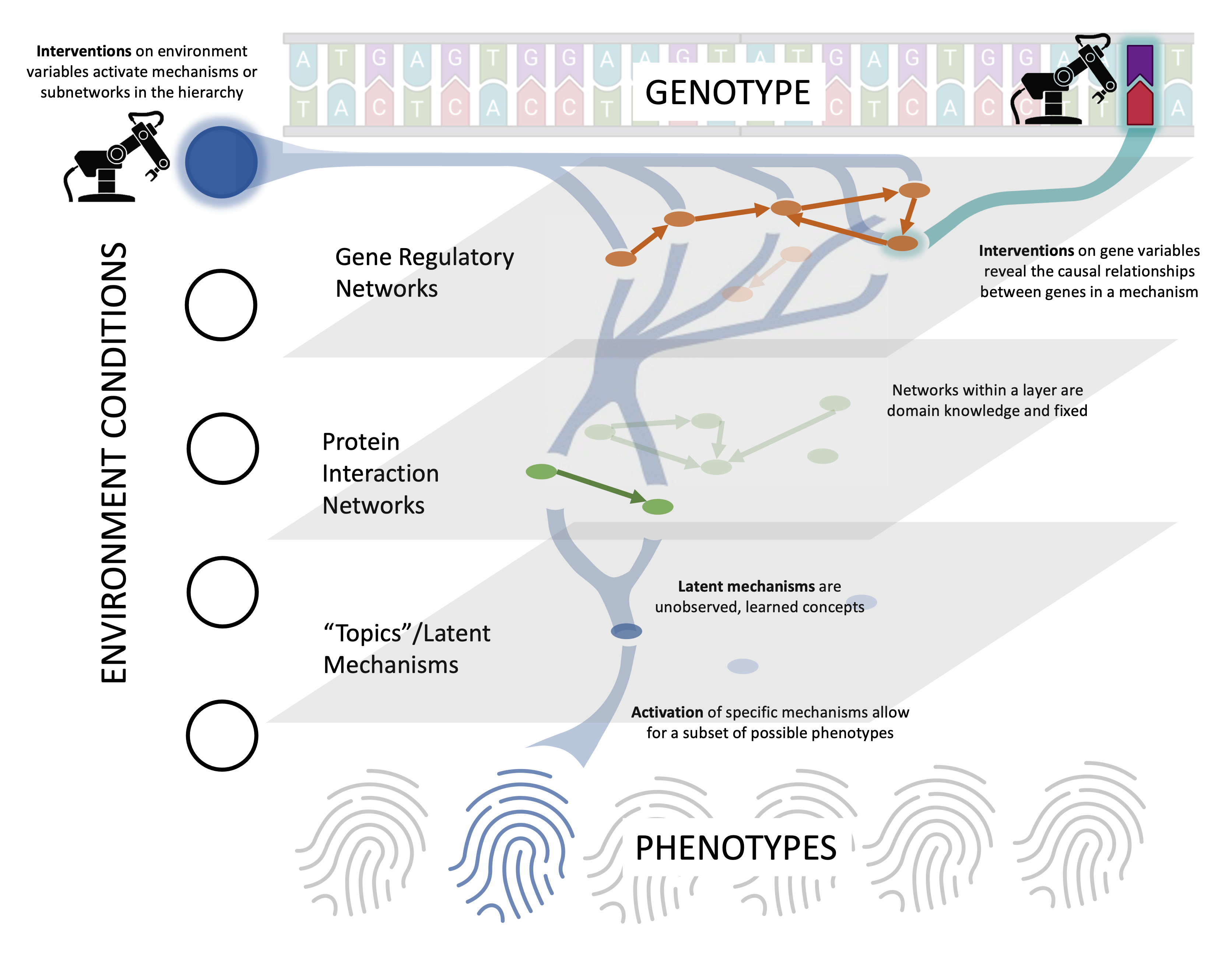}
\caption{The hierarchy of biological networks which is partially known. Levels in this hierarchy can be represented as causal networks. Interventions activate mechanisms of action and reveal causal relationships within a mechanism. The space of possible interventions is large, motivating the need for autonomous design loops like AI driven experimental design and robotic laboratories. In theory, recovery of all networks allows for full genotype to phenotype mapping. However, in practice it is impossible to fully capture the data distribution and control for confounding variables. This motivates representing mechanisms of action as topic latent variables that can be learned via topic modeling. }
\label{fig:hierarchy}
\end{figure*}

 A biological network describes causal interactions between variables in a biological system. These variables can be genes, transcription factors, proteins and metabolites. Interactions between these variables, along with external environmental factors, maps the genome of a species (or genotype) to an observable trait (or phenotype) (See \autoref{fig:hierarchy}). Predicting phenotypes from genotypes is one of the core challenges in systems biology (\cite{pigliucci2010genotype}, \cite{ritchie2015methods}, \cite{lewis2012constraining}). Genotype-phenotype models are useful for predicting cell function, disease, drug resistance, and many other problems related to biology and public health.  In this paper we focus on biological network recovery of the first layer in the interaction hierarchy (the gene regulatory network) -- which we name a ``genome-scale network'' since the number of variables corresponds to the number of genes in a genome. Recovery of genome-scale networks is important for understanding drivers for phenotypic changes and for identifying new drug targets.  Moreover, a better understanding of genome-scale networks enables more accurate downstream phenotype prediction models including those that use machine learning (See latent topics in \autoref{fig:hierarchy}). 


Biological networks are inferred from experimental data. Recovery of biological networks is a reverse engineering problem: given experimental data, what is the network that gives rise to the data? Existing methods for network recovery fall into two primary categories: pairwise information theoretic methods, and graphical model methods. The advantages of the former are that they are embarrassingly parallel and have been shown to be effective on large-scale biological datasets (\cite{faith2007large}, \cite{lachmann2016aracne}, \cite{margolin2006aracne}). The main disadvantage of these methods is that they can only learn from one data distribution -- e.g. they cannot incorporate new experimental data after an intervention has been applied to a system. The second category of network recovery methods are graphical models, often called structure learners. The advantages of these are that they have an immediate causal interpretation (the direction of an edge implies cause and effect) and as a result a suite of methods have been built to jointly learn from observational and interventional data distributions (\cite{hauser2012characterization}, \cite{wang2017permutation}, \cite{tong2001active}) -- we call these ``joint learners''. This ability to jointly learn from different distributions takes a statistical model (e.g a Bayesian Network) and converts it into a causal model (e.g a Causal Bayesian Network). The disadvantages of graphical models are that they do not scale well with the number of nodes in the graph (the graph search space is combinatorial), and joint learners do not have parallel implementations. This means that joint structure learners have not been evaluated on datasets at scale -- at most we observed evaluations on 500 node networks. Given that the genomes of most species are on the order of $10^3$-$10^4$ genes, there is a need to scale up these joint structure learners to larger networks. In this paper we introduce a new hybrid structure learner, named SP-GIES\footnote{Code is available at https://github.com/shahashka/SP-GIES}, that leverages the advantages of both categories of methods.

A causal model is preferable to a statistical model because directed causal edges in a network allow us to identify pathways, or mechanisms of action, from genotype to phenotype. In applications of biology and medicine, researchers seek to model both functional outcomes and the mechanisms leading to outcomes so that these can be understood and manipulated through therapeutics. Given this priority, there is a demand for causal models in this application space. This motivates the need to incorporate data from interventions on the system. We estimate the space of possible interventions in the environment and gene spaces to be $10^4$-$10^5$ depending on the species under investigation. A brute force sampling of this space of interventions is experimentally expensive and wasteful. Instead, there is a need to design feedback loops that choose advantageous interventions based on domain knowledge and learned knowledge. Work in optimal experimental design (OED) uses expected gain to choose interventions on variables in Bayesian Networks to drive causal discovery (\cite{tong2001active}, \cite{hauser2014two}, \cite{agrawal2019abcd}, \cite{ghassami2018budgeted}). This paradigm allows us to reason about how new inferred knowledge can connect to our existing domain knowledge, and provide a path forward for integrating AI into science domains.

The contributions of this work are as follows:
\begin{enumerate} 
    \item The implementation of a new hybrid structure learning method named Skeleton-Primed Greedy Interventional Equivalence Search (SP-GIES), based on the existing method GIES, that jointly learns from observational and interventional data. SP-GIES achieves better network recovery accuracy and a faster time to solution on large-scale networks compared to existing methods. 
    \item The application of SP-GIES to an optimal experimental design feedback loop that chooses optimal interventions on genes. 
    \item An analysis of the complexity of these methods and discussion of future directions, including unified libraries of causal algorithms, in order to achieve genome-scale network recovery for genotype to phenotype mapping with interventions.
\end{enumerate}

\section{Preliminaries}
\label{sec:prelim}

\begin{figure*}[!ht]
\centering
\includegraphics[width=15cm]{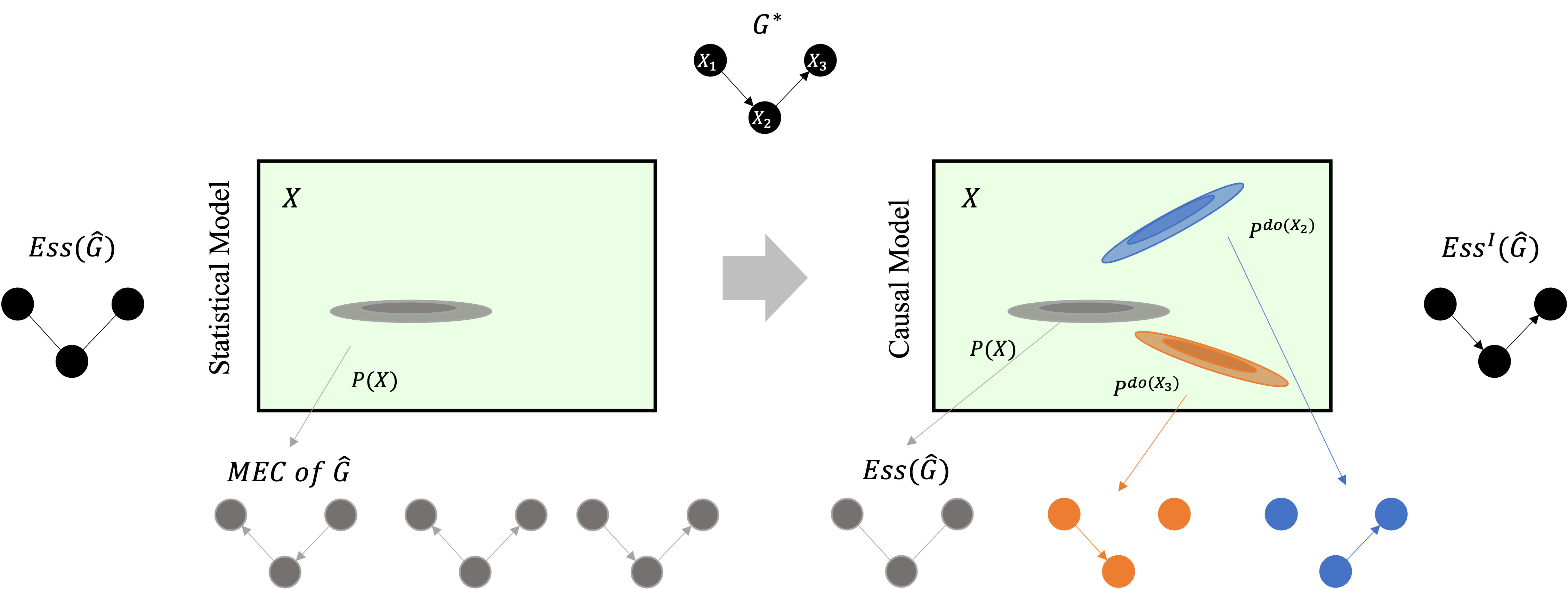}
\caption{How to convert from purely statistical model (such as a Bayesian Network) to causal model (such as a Causal Bayesian Network) with interventional data. By modeling interventional distributions, the interventional essential graph $Ess^I(G)$ is closer to the true underlying causal graph compared to the essential graph $Ess(G)$. This figure is adapted from \citet{DBLP:journals/corr/abs-2102-11107}}
\label{fig:stat_to_causal}
\end{figure*}
\subsection{Causal Bayesian Networks}
Let $G=(V,E)$ be an acyclic graph defined by a set of vertices $V$ and directed edges $E$. The vertices of the graph represent random variables $X_1...X_p$. Under the Markov Assumption for Bayesian Networks, each variable $X_i$ is conditionally independent of its non-descendants given its parents. The joint distribution of a Bayesian network factorizes as $P(\textbf{X}) = \prod_{i=1}^{p} P(X_i|Pa(X_i))$ \citep{Spirtes2000}. The following definitions describe important properties of Bayesian Networks that are relevant for structure learning. 

\begin{definition}[Markov Equivalence Class (MEC)]
The Markov Equivalence class (MEC) of a Bayesian Network $G$ consists of all directed acyclic graphs (DAGs) that share the same conditional independence relationships. 
\end{definition}

\begin{definition}[Essential Graph (Ess)]
An essential graph Ess(G) is a partially oriented graph that uniquely represents the MEC of a DAG. Directed arrows only exist on edges consistent across the equivalence class and all other edges are left undirected \citep{andersson1997characterization}.
\end{definition}

  A Causal Bayesian Network is a Bayesian Network where edges represent causal and effect relationships \citep{pearl1995bayesian}. For example if a directed edge exists from $X_i$ to $X_j$, this means that no unobserved confounding variables are responsible for the their correlation and $X_i$ is a direct cause of $X_j$. 

\subsection{Structure Learning with Observational Data}
Constructing the graph structure from a set of instances (sampled values for each node $X_i$) is called structure learning. The following assumptions are made for learning a graph $\hat{G}$ from data. 
\begin{enumerate}
    \item \textbf{Causal sufficiency} - All random variables are observed, i.e. there are no hidden variables
    \item \textbf{Causal Markov Assumption} - The data is generated from an underlying Bayesian Network $(G^*, \theta^*)$ over a set of random variables $X$
    \item \textbf{Faithfulness Assumption} - The distribution $P^*$ over $X$ induced by $(G^*, \theta^*)$ satisfies no independencies beyond those implied by the structure of $G^*$
\end{enumerate}
 We are given a data set $D = \{X_1...X_p\}$ of $N$ samples from $P^*$ -- this data is assumed to be independent and identically distributed (iid). The task is to learn a model $\hat{M} = (\hat{G}, \hat{\theta})$ that defines a distribution $\hat{P}$ that best fits true distribution $P^*$ \citep{koller2009probabilistic}.

All graphs in the same MEC give rise to the same observational distribution. Therefore the best we can do with only observational data is recover the true graph's MEC. Several graphical model based methods learn the structure from observational data only. These are split into constraint-based and score-based methods. There are also pairwise information theoretic methods that learn from only observational data. We discuss these in Section \ref{sec:related_work}.

\subsection{Structure Learning with Interventional Data} 
In addition to structure learning on observational data, we want to incorporate interventional datasets into our learning procedures. Intervening on a set of random variables $I\subseteq X$ removes incoming edges to the random variables $I$, and sets the joint distribution to a new interventional distribution  $P^{do(X_i=x)}$. Here, we are setting node $X_i$ to a value $x$. In this paper, we assume a ``hard'' interventions, where a node is set to a constant value, because this mimics the types of interventions we are able to perform in biology applications -- e.g. gene knockout experiments. This is contrast to a ``soft'' intervention which is samples $X_i=x$ from a distribution.

\begin{definition}[Interventional Distribution]
The joint distribution after a hard intervention is: \\
\begin{equation*}
\label{eq:inter_dist}
P^{do(X_i=x)} = \begin{cases}
\prod_{j \neq i}^{p} P(X_j|Pa(X_j) & if X_i=x\\
0 & otherwise
\end{cases}
\end{equation*}
\end{definition}

Jointly learning on observational and interventional data allows us to correctly isolate causal relationships and orient edges in the graph.
An estimated $Ess(G)$ from observational data can be further refined into an  I-essential graph ($Ess^I(G)$). This is a partially oriented DAG that represents an interventional Markov Equivalence class (I-MEC) (See \autoref{fig:stat_to_causal}). There are several existing methods that jointly estimate structure given observational and interventional data. See \citet{vowels2021d} for a full list of these structure learners. 

\begin{table*}[!ht]
\caption{Properties of all the learners covered in this paper. Methods are split row-wise by the nature of the learner. Data type refers to observational and/or interventional data capability of the learners. Evaluation dataset lists the benchmarks used for evaluation for each paper. Finally, the table also lists the maximum number of nodes the algorithm was evaluated on and the worst case complexity of the algorithm as reported in the corresponding papers. $p$ is the number of nodes, $n$ is the number of samples. $k$ is the maximal degree of any node in the graph. Note that for joint learners the average case complexity is not exponential, however exact calculations depend on clique size and network topology. }
\scalebox{0.8}{
\begin{tabular}{| cl||c|c|c|c|c|c|c|c|  }
\hline
&& \multicolumn{2}{c|}{Data Type}&\multicolumn{3}{c|}{Evaluation dataset} &\multicolumn{2}{c|}{Scaling}\\ 
\hline
&Paper & Observ. & Interv.  & Random & DREAM & Large-Scale & Max \# of nodes & Worst case runtime\\
\hline\hline
&\textit{CLR} \cite{faith2007large} & \checkmark & &  & &  \checkmark & 4,345 & $O(p^3n^3)$ \\
Pairwise Info. Theoretic &\textit{ARACNE-AP} \cite{lachmann2016aracne} & \checkmark & & &  &  \checkmark & 1,331 & $O(p^3 + p^2n^2)$\\
&\textit{Pinna et. al} \cite{pinna2010knockouts} &  & \checkmark& \checkmark & \checkmark& & 100 & $O(p^3+np)$\\

\hline
&\textit{PC} \cite{zarebavani2019cupc},\cite{le2016fast}, \cite{madsen2017parallel}& \checkmark &  & \checkmark & \checkmark & \checkmark &5,361 & $O(p^{k+2})$ \\
Graphical Models &\textit{GIES} \cite{hauser2012characterization} &\checkmark & \checkmark & \checkmark& \checkmark &  & 500 & $O(2^p)$\\
&\textit{IGSP} \cite{wang2017permutation} &\checkmark & \checkmark & &   & & 24 & $O(2^p)$\\
&\textit{MCMC Mallows} \cite{rau2013joint} &\checkmark & \checkmark &   & \checkmark& & 10 & $O(2^p)$\\ 
\hline
Hybrid & \textit{SP-GIES} [this paper] & \checkmark & \checkmark & \checkmark & \checkmark &\checkmark & 2,000 & $O(2^p)$ \\
\hline
\end{tabular}
}
\label{tab:refsummary}
\end{table*}

\begin{table*}[!ht]
\caption{Properties of the OED methods covered in the paper. OED criteria refers to the utility function used for selection of next experiments. Note that the the complexity listed here corresponds only to choosing the next intervention or sets of interventions. Each algorithm here also has the cost of sampling from the posterior distribution which is equivalent to the complexity of the learners in \autoref{tab:refsummary}}
\begin{tabular}{| c||c|c|c|c|c|c|c|c|  }
\hline
& \multicolumn{2}{c|}{OED criteria} &\multicolumn{3}{c|}{Evaluation dataset} &\multicolumn{2}{c|}{Scaling}\\ 
\hline
Paper  & Info. theory & Edge orient.  & Random & DREAM & Large-Scale & Max \# of nodes & Worst case runtime\\
\hline \hline
\citet{ness2017bayesian}  & &\checkmark &   & \checkmark& & 17 & $O(|E|*p!)$\\
\citet{hauser2014two}  & &\checkmark &   & & &40 & polynomial in $p$\\
\citet{tong2001active} &\checkmark &  &  & & & 12 & $O(pn)$\\
\textit{ABCD} \citep{agrawal2019abcd} &\checkmark &  & \checkmark & \checkmark& & 10 & $O(p^2n)$\\
\textit{BED} \citep{ghassami2018budgeted} &\checkmark &  & \checkmark & \checkmark& & 100 & polynomial in $p$\\
\hline
\end{tabular}
\label{tab:refsummary_oed}
\end{table*}

\subsection{Optimal Experimental Design}
 We have access to new interventional data by sampling the system, but we wish prioritize the interventional experiments that will recover the true causal graph the fastest. 
This is done via maximization of an expected utility function, $U$, over the set of potential interventions: $ \hat{I} = \argmax_{I}\mathbb{E}_{G|D}[U]$
where $U$ is a utility function like mutual information, number of oriented edges, etc. 

Our goal is to recover the underlying Causal Bayesian Network given a mix of interventional and observational samples, and some prior information about the causal edges in the graph. We cast this as a Bayesian Inference problem over the space of DAGs. We have a prior $P(G)$ which encodes any prior structural knowledge about the underlying DAG. Applying Bayes Theorem gives us the posterior distribution $P(G|D) \propto P(D|G)P(G) $. The likelihood is $ P(D|G) = \int_\theta P(D,\theta|G)d\theta = \int_\theta P(D|\theta,G)P(\theta|G)d\theta$, where we have marginalized out the parameters of the graph \citep{tong2001active}. $D$ is a mix of observational and interventional data. The posterior distribution is sampled via the joint structure learners described in the previous section. After sampling new data from the chosen intervention using OED, the new data is concatenated to the existing dataset and the posterior is re-sampled using the structure learners.

\section{Related Work}
\label{sec:related_work}
\autoref{tab:refsummary} and \autoref{tab:refsummary_oed} provide summary of the related work in biological network recovery and optimal experimental design respectively. 
\subsection{Structure Learning With
Interventional Data for Biology
Network Recovery}

Pairwise information theoretic learners use calculated metrics to estimate the dependence of two variables. A threshold is applied to the metric to obtain a network representation of the system. \citet{pinna2010knockouts} use a deviation matrix (difference between intervened samples and steady state samples) to estimate an initial network. This algorithm won the DREAM4 network recovery challenge and was the state-of-the-art for gene regulatory network reconstruction at the time. \citet{faith2007large} introduce CLR (context likelihood of relatedness) which calculates the pairwise B-spline mutual information between genes in the E. coli K-12 gene regulatory network. CLR is implemented in MATLAB, with a fast parallel kernel for calculating the pairwise mutual information. \citet{margolin2006aracne} developed ARACNE (Algorithm for the Reconstruction of Accurate Cellular Networks) which is a similar pairwise mutual information based network recovery algorithm. ARACNE additionally employs a DPI (data processing inequality) algorithm that removes indirect interactions. ARACNE was more recently upgraded to ARACNE-AP  \citep{lachmann2016aracne}, which is implemented in Java, to handle adaptive partitioning to calculate the mutual information -- this achieved a 200x computational performance improvement.  

Graphical model based learners reveal a much finer structure than pairwise methods because they are based on Bayesian Networks. The PC algorithm is a constraint-based structure learner that only handles observational data. PC learns a graph by performing conditional independence testing on pairs of nodes conditioned on other nodes. Since its introduction, many parallel implementations and optimizations of the PC algorithm have been added (\cite{le2016fast}, \cite{madsen2017parallel}) including one that uses GPUs (\cite{zarebavani2019cupc}). Existing structure learners that jointly learn from observational and interventional data (joint learners) are either score-based or hybrid score/constraint-based methods. \citet{rau2013joint} propose an algorithm called MCMC Mallows based on Markov Chain Monte Carlo sampling over the space of potential Gaussian DAGs and optimizes the best scoring DAG. MCMC Mallows was evaluated on the DREAM4 networks, and scored better than \citet{pinna2010knockouts} on two of the networks. \citet{hauser2012characterization} proposed Greedy Interventional Equivalence Search (GIES) which is a greedy score-based approach that is an extension of the original GES learner. The worst case complexity of GIES is exponential in $p$ (the number of nodes in the graph), however the authors note and show empirically that GIES is more efficient than this worst case. GIES was evaluated on DREAM4 and achieved scores in the top third of all participants. Intervention Greedy Sparsest Permutation (IGSP, \citet{wang2017permutation}) is a hybrid score/constraint, and non-parametric extension of the GSP learner. IGSP associates a DAG to every permutation of random variables and greedily updates the DAG by transposing elements of the permutation. IGSP performed comparably to GIES on protein signaling data \citep{sachs2005causal} and better than GIES on a real gene expression dataset with 24 genes. The worst case complexity of IGSP is also exponential in $p$. IGSP and GIES can be implemented with polynomial complexity by restricting the maximum degree of each node in the graph, however, for large scale networks with hubs (nodes with high connectivity) the maximum degree may be hard to determine. We show the empirical scaling of these joint learners against our proposed learner, SP-GIES, in Section \ref{sec:method}.

\subsection{Optimal Experimental Design for Biology Network Recovery}
\label{sec:oed}
One way to choose the optimal intervention is to choose the set of interventions that maximizes the mutual information or leads to the greatest decrease in entropy.
\citet{tong2001active} sample a set of orderings from the distribution over graphs and parameters. They then use this set of orderings to compute entropy terms and select the intervention with the lowest expected posterior loss. \citet{agrawal2019abcd} extend the work of \citet{tong2001active} with the ABCD strategy --  a greedy implementation of batched experimental design. They use a utility function based on the expected entropy decrease of an intervention. This requires calculating expectations over the graph and parameter spaces, however, they are able to make a tractable algorithm by using bootstrapping and weighted importance sampling. 

Another way to choose the optimal intervention is to choose the intervention that leads to the maximal number of oriented edges. \citet{ness2017bayesian} use optimal experimental design to recover protein signaling networks \citep{sachs2005causal}. They use a utility function based on the expected information gain of an intervention given the observational MEC and other interventions in the batch. This algorithm, however, has factorial dependence on batch size. \citet{ghassami2018budgeted} use the expected number of oriented edges of an essential graph as the utility function. The essential graph of the ground truth network is first estimated using a constraint based method like the PC algorithm. \citet{hauser2014two} similarly propose a utility function based on the number of oriented edges of a skeleton graph. 

\section{Method}
\label{sec:method}
Here we describe our hybrid algorithm Skeleton-Primed Greedy Interventional Equivalence Search (SP-GIES). A skeleton is an undirected graph with edges that correspond to edges in a DAG. The algorithm is a simple sequential use of an observational learner to estimate a skeleton, and the joint graphical model structure learner GIES to orient edges. The two step algorithm is as follows:
\begin{enumerate}
    \item Use ARACNE, CLR or PC to generate a skeleton with only observational samples
    \item Use the output of (1) to restrict the possible edge set for the GIES learner. Jointly learn from observational and interventional data using GIES. 
\end{enumerate}
If the  PC algorithm is used then the input to (2) is an Ess(G) which is represented by a partially oriented DAG or PDAG.  We chose GIES for (2) since it has an open source implementation that is part of the widely used \texttt{pcalg} library in R. For scaling studies, we chose to use the R implementation of PC algorithm. This is because CLR and ARACNE are implemented in MATLAB and Java respectively. Since GIES only has an R implementation, we chose to use the R implementation of PC for our scaling studies. 

\begin{figure}[!ht]
\includegraphics[width=8cm]{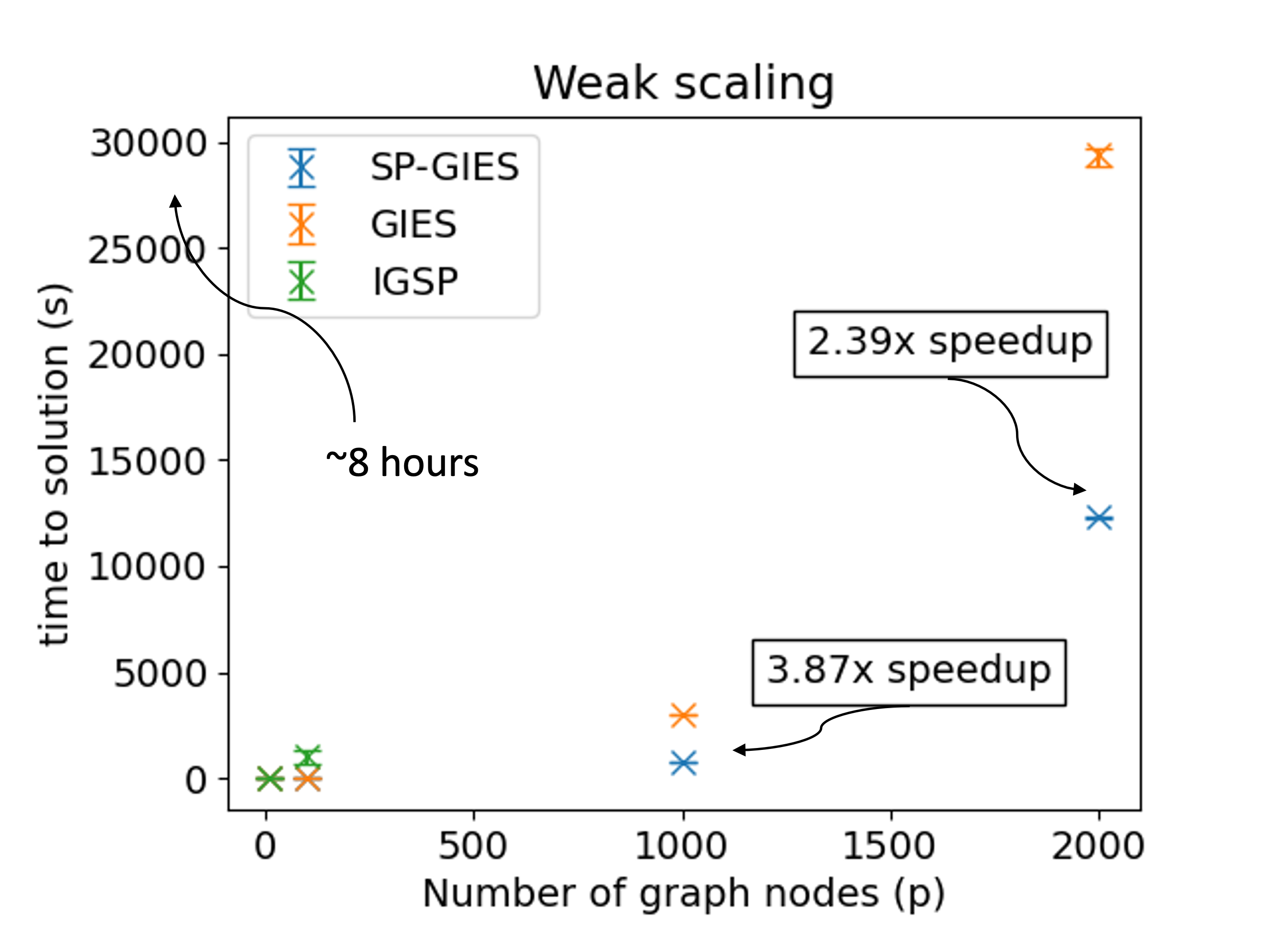}
\caption{
A weak scaling study comparing the SP-GIES, GIES and IGSP algorithms. The number of samples is fixed to $n$=1,000. Data points are averaged over 3 runs. The study was performed on a 2.8 Ghz AMD EPYC Milan 7543P 32 core CPU and a NVIDIA A100 GPU. The IGSP 1,000 and 2,000 node runs exceeded our quota and is not plotted here.}
\label{fig:scaling}
\end{figure}

\begin{figure}[!ht]
\includegraphics[width=8cm]{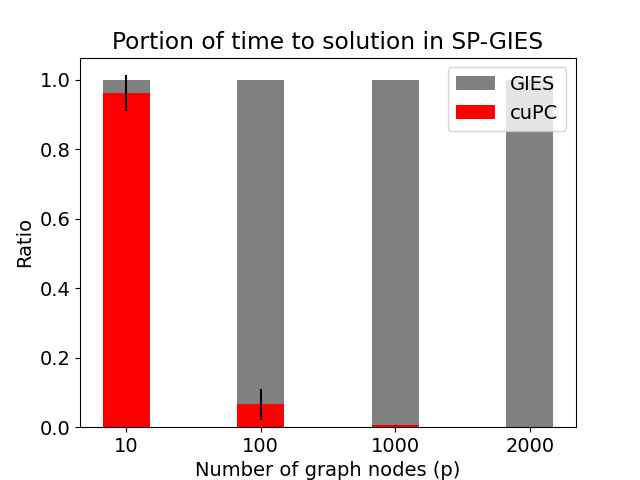}
\caption{ Fraction of runtime for each step of the SP-GIES algorithm. Step (1) is cuPC and Step (2) is GIES. We see that as the graph size increases, the bottleneck is the GIES step. }
\label{fig:bottleneck}
\end{figure}

\begin{table*}[!ht]
\caption{Evaluation of learners on  random networks of size 10. Three different types of random networks were generated: Erd\"{o}s Renyi \cite{erdHos1960evolution}, Scale free \cite{barabasi2003scale}, and Small world \cite{watts1998collective}. $p$ and $k$ refer to parameters used to generate the graphs, not the number of nodes and degree as used previously. Results are averaged over 30 random graphs, each learned with 100 data samples. The superscript on the algorithms refers to the data type used (O for observational, OI for mixed observational and interventional). For SP-GIES, ARACNE, CLR, and then PC were used for skeleton estimation for  Erd\"{o}s Renyi, Scale Free and Small World respectively since these were the best performers. }
\label{tab:randomsl}
\begin{tabular}{ |c||c|c|c||c|c|c||c|c|c|  }
 \hline
 \multicolumn{10}{|c|}{ Random Networks size 10} \\
 \hline\hline
Algorithm& \multicolumn{3}{c|}{Erd\"{o}s Renyi $(p=0.5)$}& \multicolumn{3}{c|}{Scale Free $(k=2)$} &\multicolumn{3}{c|}{Small World $(p=0.5, k=2)$} \\
\hline
   & SHD    &SID&   AUC-PR&SHD    &SID&   AUC-PR & SHD    &SID&   AUC-PR\\
   
PC\textsuperscript{O} & 22.77 & 70.73 & 0.35 & 11.27 & 61.00 & 0.62 & 8.23 & 36.33 & 0.59 \\
CLR\textsuperscript{O} &22.67 &  71.43 &  0.34 & 17.27 & 68.30  & 0.35 & 10.83 & 51.17 & 0.43 \\
ARACNE-AP\textsuperscript{O} & 21.83 & 79.00 & 0.36 & 16.63 & 76.70 &  0.37 & 10.10 &52.80  & 0.43 \\
GES\textsuperscript{O}  & 23.27 & 53.34  & 0.47&  \textbf{11.20} &     \textbf{28.57} &  \textbf{0.69} & 10.23 & \textbf{17.27} & 0.67\\
GIES\textsuperscript{OI} & 23.93 & \textbf{39.23} & 0.59 & 14.13 & 36.43 & 0.68 & 21.30 & 17.70    &    0.55\\
Pinna\textsuperscript{OI} & 22.67 & 57.60 &   0.26 & 18.00 &     56.93 & 0.28 & 13.97 & 29.73 & 0.13\\
IGSP\textsuperscript{OI} & \textbf{20.00} & 67.27 & 0.38 & 15.00 & 53.23 & 0.41 & 6.73 & 22.43 & 0.55 \\
SP-GIES\textsuperscript{OI} & 24.00 &39.87&  0.59& 11.34 & 48.53    &    0.64 & \textbf{5.90} & 26.53 &  \textbf{0.68} \\

NULL & 21.00 & 66.83 &  \textbf{0.61} & 16.00 &  61.30  & 0.58 & 10.00 & 30.00 &  0.55 \\
 \hline
\end{tabular}
\end{table*}

While (1) is easily parallelizable on multiple processors or deployable to a GPU, the bottleneck of the computation is in the GIES step (see \autoref{fig:bottleneck}). Still in \autoref{fig:scaling}, we see that SP-GIES is able to reach a faster time to solution than GIES. To understand why this is, we investigated the GIES algorithm. GIES is a greedy score-based structure learner. The score function is the Bayesian Information Criteria which is based on the maximum likelihood estimate of the graph given the current data. Starting from an empty essential graph, in the forward phase of the algorithm an edge is added to $Ess(G)$ such that the new essential graph has the maximal score. The algorithm iterates through all possible edges until it finds the corresponding essential graph with the highest score. The backward phase is similar, except an edge is removed instead of added. By restricting the set of possible edges to the GIES algorithm by first estimating the skeleton, we narrow the size of the search space for both phases -- this is what results in the speedup. Moreover, since we use a fast GPU implementation of the PC algorithm (\citet{zarebavani2019cupc}), the overhead of estimating the skeleton first is minimal. 


\begin{table*}[!ht]
\caption{Evaluation of learners on the DREAM4 networks of size 10. The dataset contains 11 observational samples and 10 interventional samples. Interventional samples are gene-knockout experiments. For SP-GIES, ARACNE was used for skeleton estimation, except for Network 4 which used CLR. Note that for Network 3, we used adaptive learning in the GIES subroutine (adaptive=``triples") to achieve the best performance for the SP-GIES learner. We did not see significant improvement with adaptive learning for the other networks.}
\label{tab:dream4sl}
\scalebox{0.8}{
\begin{tabular}{ |c||c|c|c||c|c|c||c|c|c||c|c|c||c|c|c|  }
 \hline
 \multicolumn{16}{|c|}{DREAM4 insilico size 10} \\
 \hline\hline
Algorithm& \multicolumn{3}{c|}{Network 1}& \multicolumn{3}{c|}{Network 2} &\multicolumn{3}{c|}{Network 3} &\multicolumn{3}{c|}{Network 4}& \multicolumn{3}{c|}{Network 5}\\
\hline
   & SHD    &SID&   AUC-PR&SHD    &SID&   AUC-PR & SHD    &SID&   AUC-PR & SHD    &SID&   AUC-PR & SHD    &SID&   AUC-PR\\
PC\textsuperscript{O} & 14 & 39 &  0.31 & 12 &  56 &     0.43 & 16 &  64  & 0.26 & 14 &  45 &  0.16 & 11 &  57 &  \textbf{0.59} \\
CLR\textsuperscript{O} & 13 & 36&  0.45 & \textbf{10} & 56 & 0.53 & 14 & 62 & 0.36 & 14 & 47 & 0.19 & 15 & 62 & 0.33\\
ARACNE-AP\textsuperscript{O} &\textbf{11} & 28&  0.56 & 12 & 56 &0.45 & 13 & 61 & 0.43 & 15 & 45 & 0.06 & 12 & 57 & 0.46\\

GES\textsuperscript{O}  & 13    &29&   0.49 & 15 & 52 & 0.30 & 17 & 64 & 0.15 & 19 & 48 & 0.10 & 11 & 51 & 0.14\\
GIES\textsuperscript{OI} &  13 & 31   & 0.51 & 12 & 49 &0.51 & 16 & \textbf{40} &0.49 & 15 & \textbf{26} & 0.34 &9 & 39 & 0.37\\

Pinna\textsuperscript{OI} &\textbf{11} & \textbf{22}&  0.49 & 11 & 56 &0.08 & 14 & 61 & 0.57 & \textbf{12} & 42 & 0.37 & 11 & 38 & 0.47\\


IGSP\textsuperscript{OI} & 13 & 27 &0.07& 13 &54 &0.08 &16 & 65 & 0.07 & 15 & 38& 0.06& 13 & 48 & 0.19 \\
SP-GIES\textsuperscript{OI} & \textbf{11} & 23 & \textbf{0.63} & \textbf{10} & \textbf{34} & \textbf{0.6} & \textbf{12} & 47 & \textbf{0.63} & \textbf{12} & 35 & 0.33 & \textbf{6} & \textbf{30} & 0.45 \\
NULL & 12 & 27 & 0.57 & 11 & 53 & 0.58 & 14 & 61 & 0.57 & \textbf{12} & 36 & \textbf{0.56} & 11 & 46 & 0.56 \\
 \hline
\end{tabular}
}
\end{table*}

 \autoref{tab:refsummary} shows that all joint learners have a worst case exponential complexity -- this is attributed to the combinatorial search space over graphs. However, the average empirical complexity is related to the topology of the network i.e, the size of the largest clique. Since the average complexity is difficult to calculate, we show in \autoref{fig:scaling} that SP-GIES reaches a faster time to solution as the size of the network increases. We ran a scaling study for SP-GIES using a similar study as \citet{hauser2012characterization} except we were able to scale up to $p$=2,000 nodes versus the $p$=500 nodes in the GIES paper. IGSP was unable to complete for the 1,000 and 2,000 node nodes within our allotted quota of 72 hours. We see that SP-GIES is able to reach a faster time to solution than both IGSP and GIES because of the restriction of the edge set discussed previously. 

\label{sec:evaluation}
\section{Datasets}
We tested SP-GIES against existing learners on datasets at various scales. 
Random networks were generated using the \texttt{networkx} Python library. Observational data was generated assuming a linear Gaussian model. We generated 100 observational samples and one interventional sample per node. The DREAM4 insilico challenge provides observational and interventional gene knockout data of gene regulatory networks synthetically generated from stochastic ODEs. The dataset contains 10 observational samples and one interventional sample per node. The ground truth networks are subnetworks of the larger E.coli gene regulatory network. The RegulonDB dataset was provided by \citet{faith2007large}. The network is the current estimated E.coli K12 gene regulatory network and contains 1146 nodes, 3179 edges and 524 real experimental samples. Of the 524 experimental samples, 35 are samples are taken from gene knockout experiments. The RegulonDB dataset also contains environmental interventions.  Existing learners currently only model gene variables, therefore any environment interventional samples were treated as observational data. Future work includes explicitly modeling environmental condition variables. 

\begin{table*}[!ht]
\caption{Evaluation of learners on large-scale networks. \citet{pinna2010knockouts} requires one interventional sample per gene, since the RegulonDB dataset does not provide this we did not evaluate Pinna on this dataset. GES\textsuperscript{O} required setting the maximum degree to 10 in order to achieve convergence. IGSP did not converge in 72 hours for these networks. }
\label{tab:genomesl}
\scalebox{1.0}{
\begin{tabular}{ |c||c|c|c||c|c|c|  }
 \hline
 \multicolumn{7}{|c|}{ Evaluation of Large Scale Networks} \\
 \hline\hline
Algorithm & \multicolumn{3}{c|}{RegulonDB 1146 nodes, 3179 edges} & \multicolumn{3}{c|}{Small world 1000 nodes, 1000 edges}
\\
\hline 
&{SHD} &{SID} &{AUC-PR} &{SHD} &{SID} &{AUC-PR}\\
\hline
ARACNE-AP\textsuperscript{O} & 3,752 & 25,979 & 0.04 & 1,000 & 3,883 & 0.50 \\
CLR\textsuperscript{O} & \textbf{3,095} & \textbf{18,069} & 0.30 & 2,620 & 116,560 & 0.09\\
PC\textsuperscript{O} & 3,963 & 23,908 & 0.01 & 467 & 2,447 & 0.82 \\

GES\textsuperscript{O} & 8,712  & 74,224 & 0.005 & 1,523 & \textbf{879} &0.88  \\
GIES\textsuperscript{OI} & 8,355 & 80,580 & 0.006 & 1,623 & 1,097 & 0.84  \\
Pinna\textsuperscript{OI}& x & x & x &   16,577 & 4,334 & 0.002 \\
SP-GIES\textsuperscript{OI} & 3,154 & 22,114 &0.10&   \textbf{341} & 975 & \textbf{0.91} \\

NULL & 3,179 & 18,294 & \textbf{0.50} & 1,000 & 3,883 & 0.50\\
 \hline
\end{tabular}
}
\end{table*}

Each learner was evaluated on three metrics: Structural Hamming Distance (SHD), Structural Interventional Distance (SID) and the AUC-PR. The SHD is the L1 error between the generated DAG and the ground truth DAG. The SID counts the incorrect interventional distributions as defined by Eq. \ref{eq:inter_dist}. The AUC-PR is the area under the precision-recall curve. All three of these metrics are common evaluations for causal discovery. However, we note that SHD and AUC-PR are biased towards empty graphs. Suppose a structure learner $A$ learns a graph $G_A$ with no edges and with $SHD_A=|E|$. Another learner $B$ learns a graph $G_B$ with $SHD_B > |E|$, but correctly learns some of the causal edges in the true graph. According to the SHD, $G_A$ is better and we may posit that $A$ is the better learner for the data. However, an argument can be made for $G_B$ because it captures more causal relationships than $G_A$. This means that a lower SHD may not always indicate a model that best fits the data distribution. A similar argument can be made for the AUC-PR. 
The drawback of the SID is the computational complexity is quadratic in the number of nodes for sparse networks, and cubic in the number of nodes for dense networks. This is shown empirically up to 50 nodes by \citet{peters2015structural}.
\section{Results}

 Results of our evaluation on random networks, DREAM4  networks, and genome-scale networks are shown in \autoref{tab:randomsl}, \autoref{tab:dream4sl}, and \autoref{tab:genomesl} respectively. The observational learners we evaluated are PC, ARACNE, and CLR. The interventional learners we evaluated are GIES, IGSP, Pinna and SP-GIES. We did not include MCMC Mallows in our evaluation because of the scalability challenges associated with Monte Carlo sampling. See \citet{rau2013joint} for an evaluation of MCMC Mallows on small DREAM4 and random datasets. As a reference, we also include the scores for a network without any dependencies (no edges), named NULL. For CLR and ARACNE, a threshold value must be chosen to filter edges. For CLR with the random and DREAM4 networks, we chose values so that approximately the top 10\% of edges were retained. For CLR with the RegulonDB dataset, we followed the work of \citet{faith2007large} and chose the threshold that resulted in 60\% precision. For ARACNE, we used the threshold associated with a p-value of $10^{-8}$ -- this is the default used by \citet{lachmann2016aracne}. 
 
 For random networks of size 10, the best performing learner depends on network type. GES with no interventional data performs the best on scale free networks. SP-GIES performs best on small world networks. Both GIES and SP-GIES perform comparably for Erd\"{o}s Renyi networks. For DREAM4, SP-GIES most frequently gets the best score across all metric types over all five networks. We see that the addition of interventional data improves the performance of algorithms that can handle both data types (namely GIES, and SP-GIES). 
 
\begin{figure}
     \centering
     \includegraphics[width=8cm]{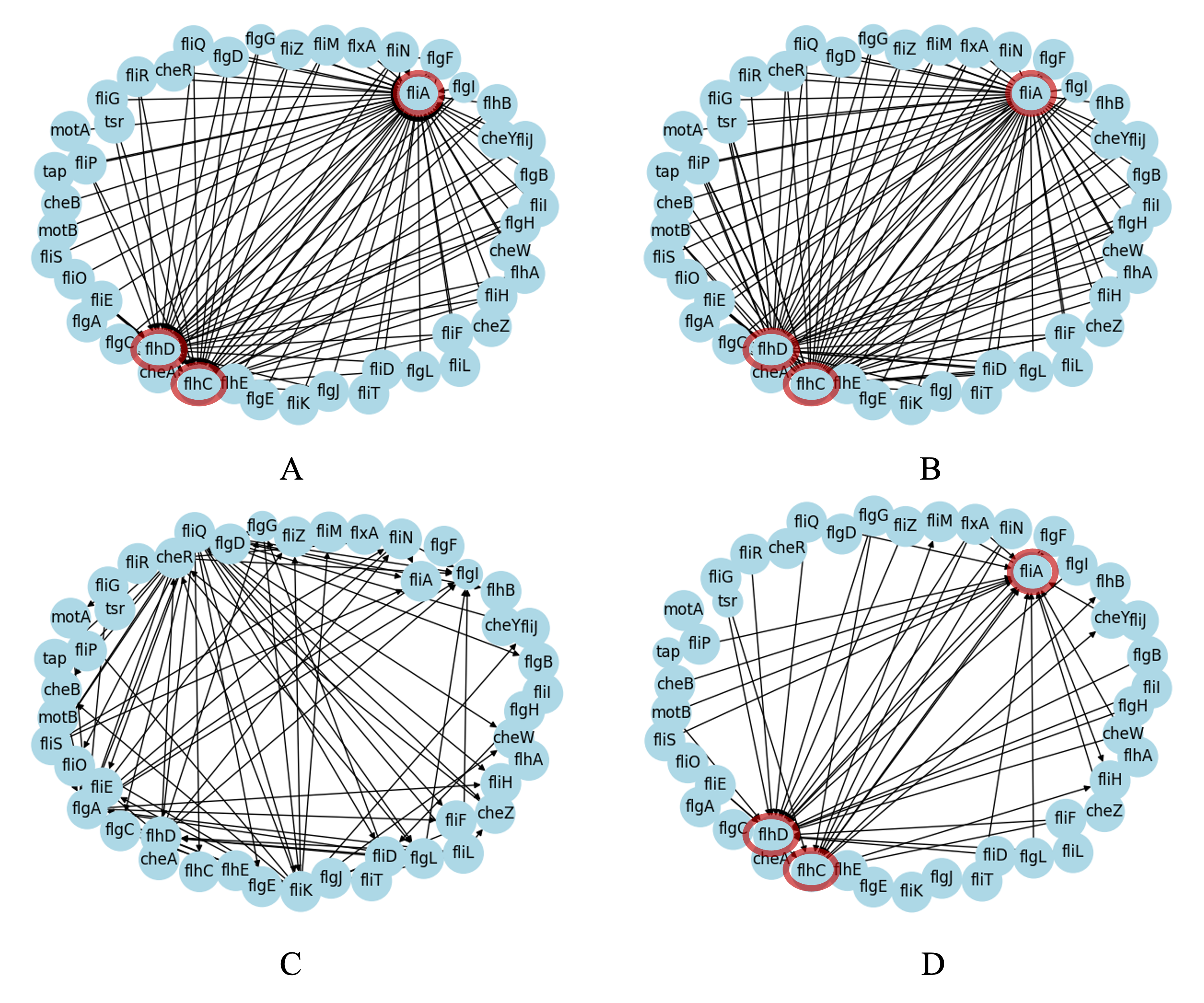}
     \caption{(A) The subnetwork of the ground truth network given by the RegulonDB dataset, which contains 3 hubs highlighted in red. (B) The estimated undirected subnetwork given by CLR, which correctly identifies the 3 hubs. The threshold was chosen so that 60\% of the ground truth network was recovered. (C) The subnetwork estimated by SP-GIES using the CLR skeleton. With real data, no hubs are detected. (D) The subnetwork estimated by SP-GIES using the CLR skeleton with synthetic data. The 3 hubs are detected, although the graph is much more sparse than the CLR estimate and ground truth network. }
     
     \label{fig:subnetworks}
 \end{figure}

 For the RegulonDB network, the best scoring method is CLR. This is because the CLR method calculates a threshold value to prune edges so that the learner achieves 60\% precision compared to the ground truth network -- this makes it an unfair comparison since the ground truth network must be known a priori. We used CLR for the skeleton estimation of SP-GIES. An unexpected result is that SP-GIES appears to worsen the initial estimate of the CLR skeleton. To understand this, we zoomed in on a subnetwork of RegulonDB that contains three hubs (highly connected nodes). \autoref{fig:subnetworks} shows that CLR correctly identifies 3 hubs in the subnetwork. However, SP-GIES removes many edges from this initial estimate and adds new edges elsewhere. As a result, no hubs are detected using SP-GIES. One possible reason is that the GIES learner uses the maximum likelihood estimate to score each candidate graph. This calculation assumes the data is sampled from a linear Gaussian distribution -- which is not the case for real datasets like RegulonDB where nonlinear effects may be present. Since mutual information can capture nonlinearity, we do not see this issue reflected in CLR. To test this theory, we generated synthetic linear Gaussian data from the RegulonDB subnetwork topology. \autoref{fig:subnetworks} shows that with synthetic data the SP-GIES can detect 3 hubs in the subnetwork. The nonlinearity of the dataset is only a partial explanation because the SP-GIES estimate with synthetic data is still further from the ground truth compared to the CLR estimate. Further analysis is needed for understanding the data regimes and network topologies within the scope of joint learners like SP-GIES.

 \begin{figure*}
\includegraphics[width=15cm]{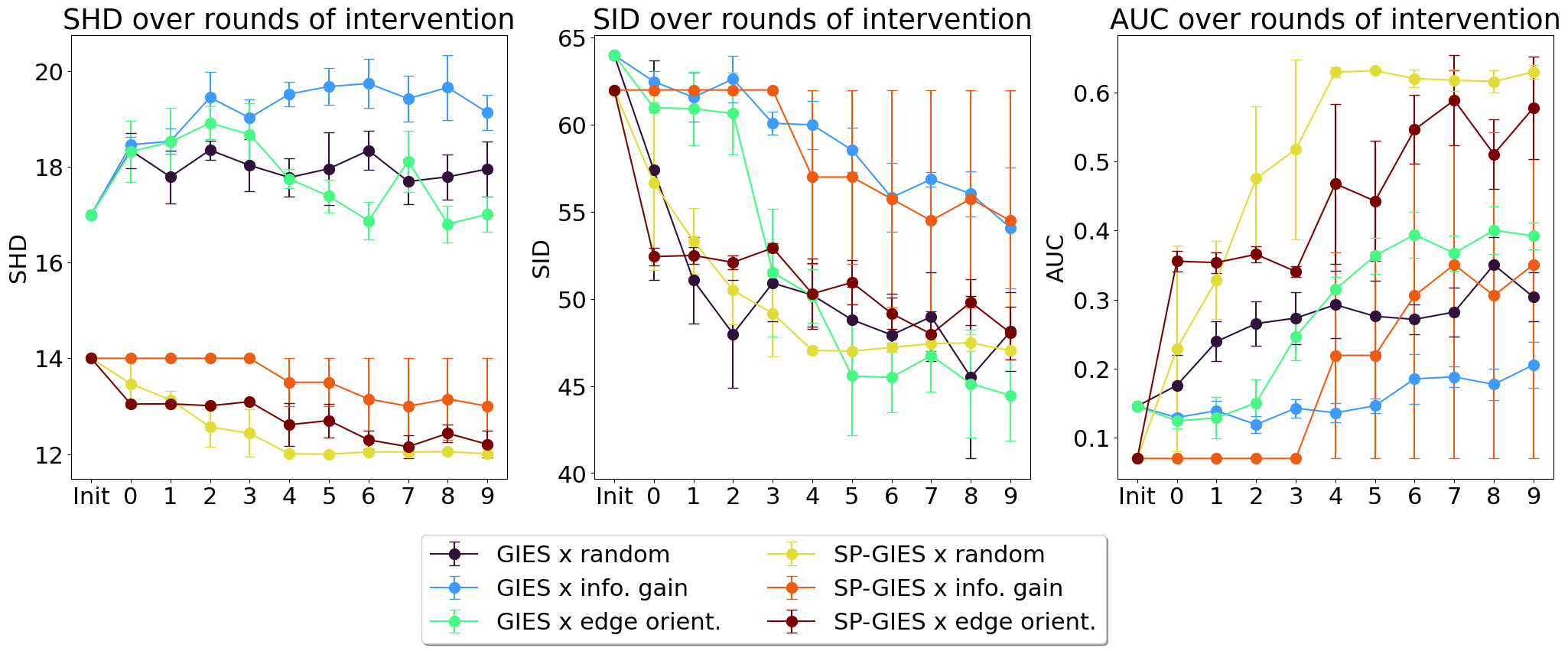}
\caption{An evaluation of OED strategies for DREAM4 10 node network \#3 over 10 rounds of intervention. At each round of intervention, a new interventional data sample is added to the current dataset corresponding to the chosen intervention. GIES and SP-GIES are joint learners. Metrics are averaged over 30 runs. }
\label{fig:dream4oed_3}
\end{figure*}
For a few of the networks we evaluated here, the NULL graph achieves the best score. Many real world networks, including gene regulatory networks, are sparse. As a result, the NULL learner often appears to perform very well since the true networks lie close to an empty network in combinatorial space. Interestingly, the NULL graph never scores best for the SID metric. This could indicate that causal network recovery methods should be evaluated with the SID rather than other metrics. However, more analysis is needed to understand the validity of this and is outside the scope of this paper.



To understand the performance gains of SP-GIES over GIES on optimal experimental design, we evaluated both algorithms with two different OED criteria on the DREAM4 insilico 10 node dataset. We limited this evaluation to small networks since the time complexity of OED includes posterior sampling and optimal gene selection for each round of intervention. We follow the framework of \cite{agrawal2019abcd}, which includes bootstrap sampling of the posterior distribution. An example result for a DREAM4 insilico network is shown in \autoref{fig:dream4oed_3}. We observe three trends here. First, the edge orientation strategy achieves better performance than the information gain strategy. Second, the SP-GIES learner boosts the initial performance of the learner, however, it easily stagnates and additional interventional samples do not continue to improve the algorithm at the same rate as the GIES based strategies. Still the SP-GIES x OED runs achieve better performance than GIES x OED, and there is some improvement in accuracy with the addition of interventional data. We speculate that stagnation may happen with the SP-GIES joint learner because by fixing the skeleton, we also restrict the search space of the learner. This may lead to the GIES step of the algorithm getting stuck in a local minima. This motivates incorporating some level of uncertainty into the skeleton, or randomly including sets of edges outside the skeleton into the search space -- we leave this to future work. Third, we see that the random strategy is comparable to the OED strategies. In other words, prioritizing certain experiments over others does not necessarily result a better estimate of the causal network. One possible explanation is that the network topology may be such that intervening on certain nodes is not advantageous compared to others. However, another reason is that the OED strategy relies on a good model (causal graph) to estimate the posterior distribution. Without a good model, the calculation of the utility function will be inaccurate and the wrong experiments may be prioritized. Although SP-GIES provides us with a better estimate of the causal graph compared to GIES, it may not be sufficient for OED. This is analogous to what we see in active learning; when the model performance is poor, a random labeling of new samples is effective for model improvement. 

\section{Discussion}
\label{sec:discussion}
Understanding genome-scale networks is important for building a causal mapping from genotype to phenotype. However, recovery of genome-scale networks from interventional datasets has not yet been realized because of the computational complexity of structure learners that jointly learn from mixed data. Existing methods like GIES, IGSP, MCMC Mallows and others are only evaluated on small networks up to 500 nodes and exhibit worst case exponential complexity. This makes subsequent optimal experimental design techniques computationally intractable since the complexity is then multiplied by the computation of choosing the optimal intervention. This is a huge issue in realizing causal discovery at scale via autonomous design loops. 

In this paper, we note that structure learners built for observational datasets have parallel implementations on multiple processors and on GPUs. Our SP-GIES learner first estimates a skeleton using these parallel implementations. This restricts the edge set for the subsequent joint structure learner and improves the time to solution. However, SP-GIES is still bottlenecked by the scaling of the joint learner GIES -- for example running network sizes of $10^4$ nodes did not complete in 24 hours. To realize recovery of genome-scale networks on the order of $10^3$-$10^4$, and to sample an interventional space on the order $10^4$-$10^5$, we must have breakthroughs in distributed parallel implementation of joint learners. Joint structure learners currently do not have parallel implementations because they are score-based learners (or hybrid constraint/score-based learners). At each step of the algorithm the candidate graph is scored, typically using a function of the likelihood $P(D|G)$ over the entire graph. As we saw in Section \ref{sec:prelim}, the joint distribution of a DAG is a product of conditional probabilities which, for each node, depends on the parent nodes. As a result, the calculation of the likelihood is sequential and typically optimization over this space is done greedily. It is not intuitive how the graph may be partitioned into subgraphs for individual compute kernels. Constraint based methods, on the other hand, use pairwise conditional independence testing to resolve edges. A naive approach to scaling is to generate a separate compute thread per calculation, however, one can group variables into blocks to further minimize computations and boost performance. This type of performance enhancement has not yet been realized by score-based learners, and this presents a roadblock for existing joint learners. 

The performance of SP-GIES on the RegulonDB dataset -- which is the most biologically relevant dataset -- suggests that future work should also incorporate nonlinearity into joint learners. One approach is to modify the GIES score function. Since closed form solutions for the maximum likelihood estimate only exist for a small set of functions, an alternative is to use a nonparametric learner like IGSP. However, IGSP does scale to the size of the RegulonDB dataset. Moreover, the IGSP learner in general appears to perform poorly on random and DREAM4 networks. We note that the use case for this algorithm is considerably different than the datasets used here. IGSP was designed for cases where there are sufficient interventional data samples present in order for hypothesis testing to be successful. For the datasets used here, only a handful of interventional samples are available for each node. Structure learning for nonlinear datasets has been studied in works such as \citet{gretton2009nonlinear}, \citet{yu2019dag}, \citet{zheng2018dags}. Recent works use neural networks by recasting the combinatorial optimization into a continuous optimization. However, these methods suffer in low data regimes. In order to realize network recovery at scale for real biological datasets, there is a need to incorporate strategies from nonlinear structure learners into joint learners. 

In evaluating existing methods and designing the SP-GIES method, we found that learners and OED strategies are implemented in a varied set of languages including Java, MATLAB, R, Python and C. This is due to the diverse nature of backgrounds interested in causal discovery of biological networks including from biology, statistics, computer science, machine learning, and representation learning. There has been a recent interest in unifying tools for causal discovery -- for example \texttt{CausalDiscoveryToolbox} \cite{kalainathan2019causal} , \texttt{Pgmpy} \cite{ankan2015pgmpy} and \texttt{CausalDAG} \cite{squires2018causaldag} are Python libraries that provide varying support for graphical and/or causal modeling. However, in the case of \texttt{CausalDiscoveryToolbox}, the library is a Python wrapper for R code, which in turn is a wrapper for C code. For \texttt{Pgmpy}, no joint interventional learners are implemented. \texttt{CausalDAG} provides the best overall framework for causal discovery with interventions, although many of the implementations are still incomplete. To realize better parallel implementations of causal algorithms and OED strategies, we should encourage collaborations with the supercomputing community. Therefore, there is a need for exclusive Python or C implementations of these models and algorithms since these languages are popular in the supercomputing community.
\section{Conclusion}
We present SP-GIES, a joint structure learner that leverages parallel observational learners to estimate a skeleton and initialize the GIES joint learner. We provide a systematic evaluation of SP-GIES against existing methods for datasets at various scales and with various metrics. We see that SP-GIES is able to provide better network recovery accuracy for larger scale networks up to 2,000 nodes and on biological networks up to 1,146 nodes. This scale of network recovery for joint learners has not been achieved to our knowledge. SP-GIES reaches a faster time to solution than existing method GIES and provides up to 3.87x speedup. We also show that SP-GIES improves subsequent OED strategies. Future work includes a distributed parallel implementation of SP-GIES, and incorporation of SP-GIES into an autonomous experimentation design loop.


\begin{acks}
Research was supported by as part of the CANDLE project by the DOE-Exascale Computing Project (17-SC-20-SC). This research used resources of the Argonne Leadership Computing Facility, which is a DOE Office of Science User Facility supported under contract DE- AC02-06CH11357 and National Institute of Allergy and Infectious Diseases, National Institutes of Health Award Number P01AI165077 (AR). This project has been funded in whole or in part with Federal funds from the National Institute of Allergy and Infectious Diseases, National Institutes of Health, Department of Health and Human Services, under Contract No. 75N93019C00076, awarded to the University of Chicago. This project has been funded in whole or in part with Federal funds from the Biological and Environmental Research program, Office of Science, Department of Energy, under FWP\# 34903, awarded to Argonne National Laboratory.
\end{acks}
\bibliographystyle{ACM-Reference-Format}
\bibliography{ref.bib}

\end{document}